\documentclass[10pt,a4paper]{article}
\usepackage[latin1]{inputenc}
\usepackage{amsmath}
\usepackage{amsfonts}
\usepackage{amssymb}
\usepackage{graphicx}
\usepackage{xcolor}
\usepackage{lineno}

\usepackage{hyperref}
\newcommand{\footremember}[2]{%
   \footnote{#2}
    \newcounter{#1}
    \setcounter{#1}{\value{footnote}}%
}
\newcommand{\footrecall}[1]{%
    \footnotemark[\value{#1}]%
} 

\usepackage{pgfplots}
\pgfplotsset{compat=newest}
\usetikzlibrary{plotmarks}
\usetikzlibrary{arrows.meta}
\usepgfplotslibrary{patchplots}
\usepackage{grffile}

\newlength\figureheight
\newlength\figurewidth

\usepackage[left=1.5cm,right=1.5cm,top=1.5cm,bottom=1.5cm]{geometry}

\author{Cynthia Michalkowski\footremember{note2}{Robert Bosch GmbH, Center for Research and Development, Robert-Bosch-Campus 1, 71272 Renningen, Germany }, Kilian Weishaupt\footremember{note1}{Institute for Modelling Hydraulic and Environmental Systems, University of Stuttgart, Pfaffenwaldring 61, 70569 Stuttgart, Germany}, Veronika Schleper\footrecall{note2}, Rainer Helmig\footrecall{note1}}
\title{Modeling of two phase flow in a hydrophobic porous medium interacting with a hydrophilic structure}
\date{January 2022}

\begin{document}
\maketitle

\begin{abstract}
Fluid flow through layered materials with different wetting behavior is observed in a wide range of applications in biological, environmental and technical systems. Therefore, it is necessary to understand the occuring transport mechanisms of the fluids at the interface between the layered constituents. Of special interest is the water transport in polymer electrolyte membrane fuel cells (PEM FC). Here, it is necessary to understand the transport mechanisms of water throughout the cell constituents especially on the cathode side, where the excess water has to be removed. This is crucial to choose optimal operating conditions and improve the overall cell performance.
Pore-scale modeling of gas diffusion layers (GDLs) and gas distributor has been established as a favorable technique to investigate the ongoing processes. Investigating the interface between the hydrophobic porous GDL and the hydrophilic gas distributor, a particular challenge is the combination and interaction of the different material structures and wetting properties at the interface and its influence on the flow.
In this paper, a modeling approach is presented which captures the influence of a hydrophilic domain on the flow in a hydrophobic porous domain at the interface between the two domains. A pore-network model is used as the basis of the developed concept which is extended to allow the modeling of mixed-wet interactions at the interface.
The functionality of the model is demonstrated using basic example configurations with one and several interface pores and it is applied to a realistic GDL representation in contact with a channel-land structured gas distributor.
\end{abstract}
\section*{Article Highlights}
pore-network model, two phase flow, hydrophobic-hydrophilic interaction, wettability
\section*{Declarations}
\subsection*{Funding}
This work was funded by the Robert Bosch GmbH and supported by the Deutsche Forschungsgemeinschaft (DFG, German Research Foundation) by funding SFB 1313, Project Number 327154368.
\subsection*{Conflicts of interest}
The authors declare that they have no conflict of interest.
\subsection*{Code availability}
All code is available online in the Dumux gitlab repository (https://git.iws.uni-stuttgart.de/dumux-pub/michalkowski2021a)
\subsection*{Availability of data and material}
All relevant data can be generated with the open source code.
\subsection*{Authors' contributions}
All authors contributed to the study conception and design. Material preparation, data collection and analysis were performed by Cynthia Michalkowski and Kilian Weishaupt. The first draft of the manuscript was written by Cynthia Michalkowski and all authors commented on previous versions of the manuscript. All authors read and approved the final manuscript.
\newpage
\section{Introduction}
\label{intro}
\begin{figure}[h!]
	\includegraphics[width=\textwidth]{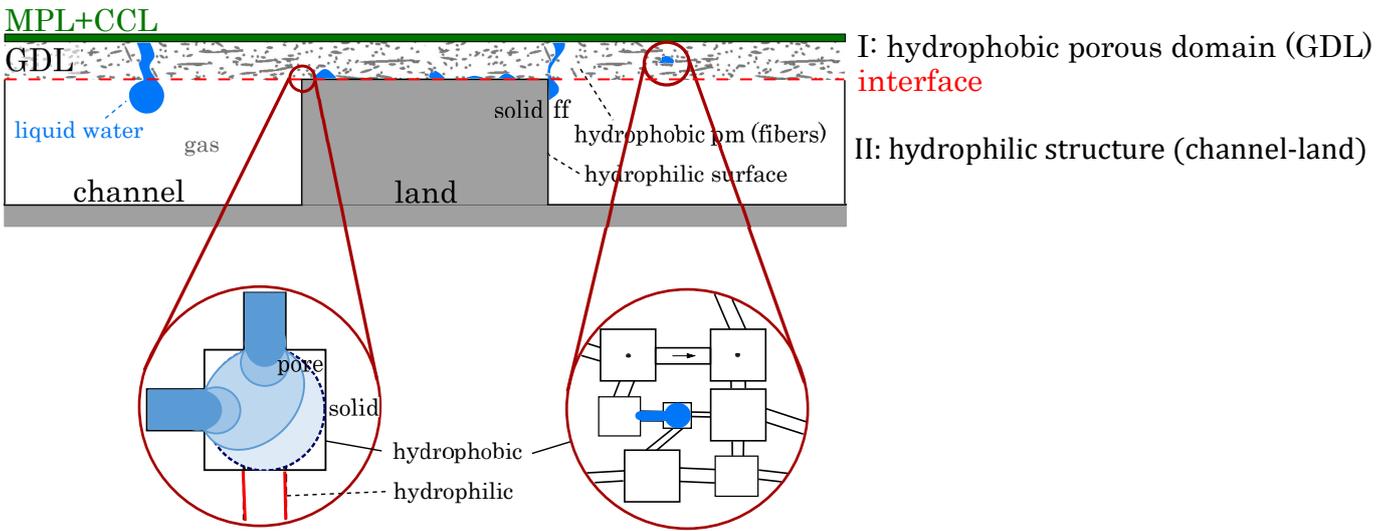}
	\caption{Cathode of a PEM fuel cell with the different interface configurations occurring between the GDL (domain I) and the gas distributor (channel-land) (domain II). Water is passing through the hydrophobic GDL, in liquid form, interacting with the free flow (ff) and the hydrophilic surface of the channel walls of the gas distributor.}
	\label{fig:CathodeConfigurations}
\end{figure}

Developing coupling concepts to combine porous medium and free flow has received attention in recent years due to the wide range of applications in biological \cite{vidotto2019hybrid}, environmental \cite{vanderborght2017heat} and technical devices \cite{torkzaban2007resolving,dahmen2014numerical,molaeimanesh2016role,zenyuk2015coupling,alink2014coupling}.\\
Of special interest is the modeling of flow in hydrophobic porous media interacting with free flow as it can be found in several technical application such as filters \cite{torkzaban2007resolving} and polymer electrolyte membrane (PEM) fuel cells \cite{molaeimanesh2016role,zenyuk2015coupling,alink2014coupling}.\\
Even though the presented concept is developed to model the fluid interactions in a PEM fuel cell, it has a wide range of applications, where similar phenomena resulting from hydrophobic-hydrophilic interactions need to be captured, such as filter materials \cite{chase2013mixed}, hygiene products \cite{benecke2003color} or protective masks \cite{ahn2010hydrogel}.
Nevertheless, in the following, we focus on discussing the concept for modeling of the processes in the cathode of a PEM fuel cell.\\
The concept presented in this work is a modeling approach for pore-scale processes at the interface between the hydrophobic gas diffusion layer (GDL) and the hydrophilic gas distributor. The GDL material is a Polytetrafluorethylen (PTFE) coated carbon fiber paper, while for the gas distributor, a channel-land structure with hydrophilic surface properties is considered. In Fig.\ref{fig:CathodeConfigurations}, the investigated setup is shown schematically (two-dimensional visualization). In this work, the interaction of water in the GDL (domain I) with the hydrophilic surface of the gas distributor (domain II) is considered (red dashed line marks the interface between the GDL and the gas distributor). The left red circle marks the domain of special interest. In the right red circle a schematic representation of the pore network of the porous GDL structure is shown.  
\\
We discuss the behavior of two-phase flow (air and water) in two domains:
\begin{itemize}
	\item Domain I represents the hydrophobic GDL. Air is the wetting phase in this domain. Drainage is the displacement of air by water (liquid).
	\item Domain II represents the hydrophilic gas distributor. Here, water is the wetting phase.
\end{itemize}
We are interested in the influence of domain II on the flow behavior in domain I, especially due to the wettability change at the interface between I and II. Locally, pores in domain I become filled with water. The interaction between I and II also results in emptying of single pores, which means the displacement of water by air in a pore.
\\
REV based models (representative elementary volume \cite{bear2013dynamics}) yield volume averaged quantities such as overall saturation in the porous material, temperature distributions or other volume averaged based quantities. However, they lack the capabilities of describing local pore-scale effects in sufficient detail for certain microscopic interface driven systems, like local water content or local compositions of fluids, which might be relevant for fuel cells. For instance, Shahraeeni et al.  \cite{shahraeeni2012coupling} showed that local, pore-scale water content patterns at the surface of a drying soil play a crucial role for the entire system, while they cannot be captured by REV-scale models. In REV-scale models local (pore-scale) wettability changes cannot be described. Pore-scale models are needed to describe local flow characteristics like preferential flow paths and the effects of pore-local wettability characteristics on the flow behavior.\\
According to \cite{lee2010steady}, invasion-percolation with capillary fingering is a dominant transport mechanism for two fluid phases flow in GDLs for common PEM fuel cells operating conditions.\\
Due to the importance of pore-scale geometric properties (pore radius up to 100 $\mu$m) in a relatively large domain of interest (several square millimeters), modeling the coupled interaction of these different materials is a special challenge. It is difficult to capture the local effects with a sufficiently fine resolution with reasonable computational effort. In recent years, several approaches have been used to capture the transport through the porous layers of a PEM fuel cell \cite{straubhaar2016pore,carrere2019liquid,zhang2018three,raeini2014direct,zenyuk2015coupling,alink2014coupling}. 
Particularly interesting for the representation of the porous structure of the GDL is the application of pore-network models \cite{straubhaar2016pore,carrere2019liquid,qin2015water}. Pore-network models offer a decent degree of accuracy on the pore-scale at comparatively low computational costs \cite{weishaupt2019efficient}. They have been used successfully for the simulation of pore-scale processes such as two-phase displacement processes \cite{thompson2002pore}, drying \cite{laurindo1998numerical} or solute transport \cite{mehmani2017minimum}, where they perform astonishingly well in comparison to more involved numerical methods, given their rather simple and efficient nature \cite{oostrom2016pore}. In general, one distinguishes between static and dynamic pore-network models \cite{joekar2012analysis}. Dynamic pore-network models capture the transient flow behavior while static networks calculate equilibrated, stationary configurations based on the applied initial and boundary conditions. For our purpose, we need a dynamic formulation to take the transient interface processes into account.\\
Weishaupt et al. \cite{weishaupt2021dynamic} present a dynamic pore-network model, which we extended to enable the modeling of hydrophobic-hydrophilic interactions of two-fluid-phases flow on the pore-scale. In the right circle in Fig.\ref{fig:CathodeConfigurations}, the representation of the GDL structure using a pore network is visualized schematically.\\
The aim of this study is to extend this pore-network approach to allow the modeling of the interaction of a hydrophilic channel-land structure (domain II) with a hydrophobic porous medium (domain I) as shown in Fig.\ref{fig:CathodeConfigurations}. We investigate the influence of mixed-wet interface processes between the hydrophobic GDL (domain I) and the hydrophilic channel-land structure (domain II) on the flow. The flow in the hydrophobic GDL is modeled on the pore-scale and we discuss the possibilities of a simplified porous representation using the extended dynamic pore-network model following Weishaupt et al. \cite{weishaupt2021dynamic}. \\
The hydrophobic-hydrophilic interaction method is based on the widely established pore-network approach following \cite{thompson2002pore} and \cite{joekar2010non} and uses the same principles defining the throat conductivities and local capillary pressure saturation relations for different phases dependent on the wettabilities.\\
The paper is organized as follows: First, the pore-network model and the formulation of the model for an interface between hydrophobic domain I and hydrophilic domain II are presented. Afterwards, the numerical model and the implementation are described. The behavior of the promoted model is investigated in applications in the numerical studies section. Finally, the numerical results are discussed and further research topics following this work are presented.
\section{Model concepts}
\label{sec:ModelConcepts}
The presented concept is developed for interactions of water flow in hydrophobic porous materials (domain I) with hydrophilic structures (domain II). It is developed for the application at the interface between a hydrophobic GDL (domain I) and a hydrophilic channel-land structure (domain II), shown in Fig.\ref{fig:CathodeConfigurations}. However, the model can be applied to any interface between a hydrophobic porous domain and an open, hydrophilic domain. Therefore, in the following, the general domain descriptions are used: we consider a hydrophobic porous domain I in contact with a hydrophilic domain II.
The flow of water (non-wetting phase) in the hydrophobic region (domain I) is influenced by the hydrophilic domain II once the water gets in contact with the hydrophilic surface. Since we are not directly interested in the water flow in the hydrophilic gas distributor (domain II) but rather on the influence of this domain on the hydrophobic region (domain I), a boundary model concept is chosen. To investigate the influence of a water attracting zone at the boundary on the flow behavior, a coupling condition is developed which represents the effects of a hydrophilic region (domain II), where water invasion would take place. This boundary condition replaces the gas distributor domain (domain II).\\
The hydrophobic domain (GDL in Fig.\ref{fig:CathodeConfigurations}, domain I) is represented by a pore-network model following the approach developed by Weishaupt et al. \cite{weishaupt2021dynamic}. The model is extended enabling the modeling of hydrophobic-hydrophilic fluid interactions at the interface, which is shown by the red dashed line in Fig.\ref{fig:CathodeConfigurations}.\\
The term water invasion describes the displacement of the gas phase by liquid water in a pore. Analogeously, emptying means the decrease of the amount of water in a pore.
\subsection{Pore-network modeling}
\label{sec:PNMBasics}
\begin{figure}
	\centering
	\includegraphics[width=0.4\textwidth]{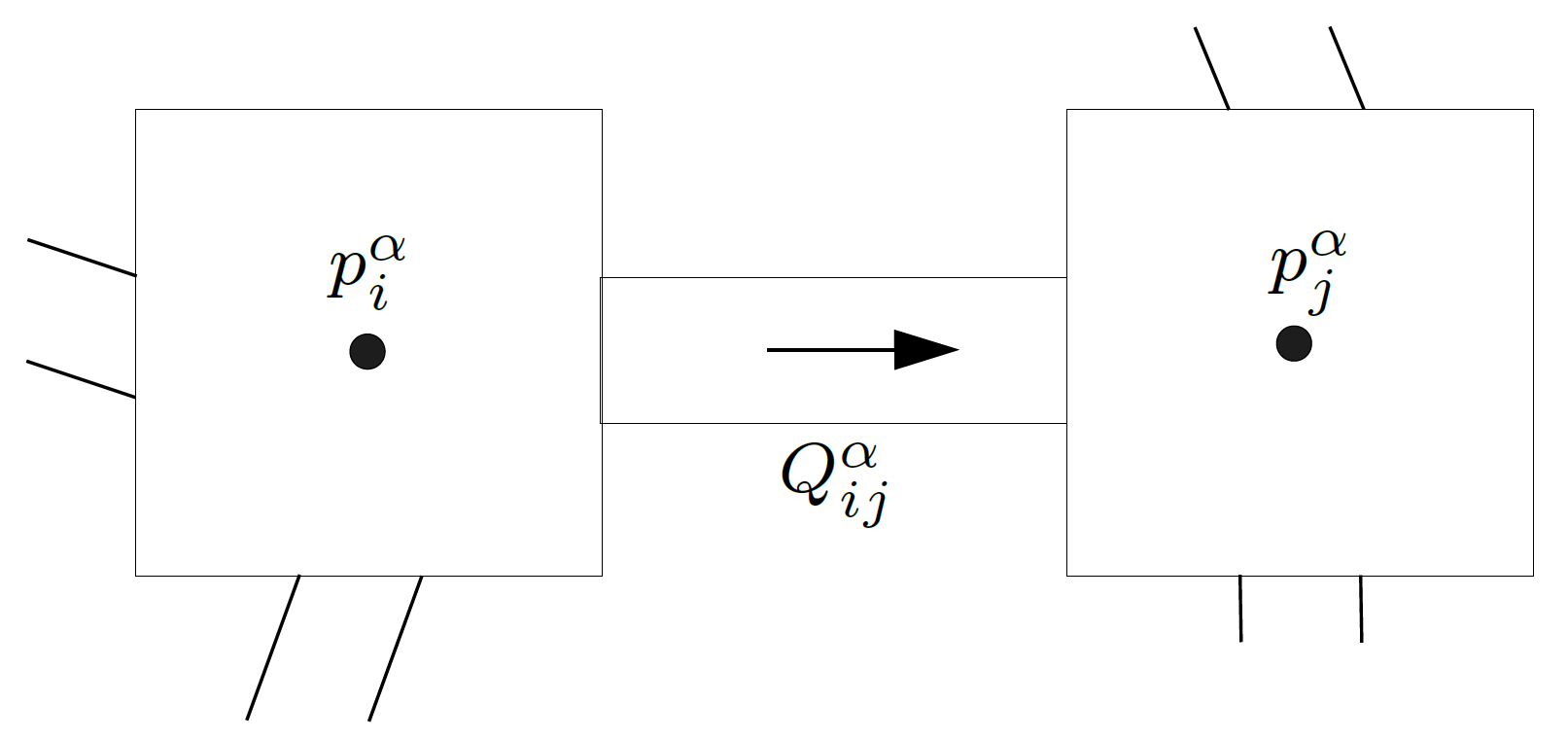}
	\caption{Flux of phase $\alpha$ between pore $i$ and $j$ with cubic pore bodies.}
	\label{fig:PNM-flux}
\end{figure}

In this paper $s$ denotes the local saturation (in a pore body), which is the ratio of the volume of one fluid phase in the pore to the volume of the pore body ($s_{w/n} = \frac{V_\mathrm{phase}^\mathrm{pore}}{V_\mathrm{pore}}$) and $S$ is the total saturation of the network describing the amount of the phase present in the complete network with respect to the overall void volume of the network ($S_{w/n} =  \frac{\sum_\mathrm{pores} V_\mathrm{phase}^\mathrm{pore}}{V_\mathrm{void}^\mathrm{newtork}}$). The sum of the saturation of all present phases equals one ($s_w + s_n = 1$ in each pore and $S_w + S_n = 1$ in the whole network).\\
The indices $w$ and $n$ refer to the wetting and the non-wetting phase, respectively. If the saturation or a fluid property refers to the water phase (liquid) independent of the wettability, the subscript $liq$ is used. For the gas phase, this holds analogously with the subscript $gas$. The terms filling and emptying of a pore refer to the water phase behavior is this work.\\
The model allocates the volume of void space of the hydrophobic porous material (domain I) to the pore bodies, which are assumed to have a negligible hydrodynamic resistance to the flow compared to the pore throats. Additionally, it is assumed that the volume of pore throats is negligible compared to the volume of pore bodies and the time required for filling a single pore throat is negligible compared to that of a pore body \cite{joekar2010non}.\\
For a more detailed introduction to the principles of pore-network modeling, it is referred, e.g., to \cite{blunt2017multiphase}.\\
The continuity equation for a pore body $i$ for each phase $\alpha$ is given by
\begin{equation}
	V_i \frac{\partial s_i^\alpha}{\partial t} \rho_i^{\alpha} = \sum_j \left(\rho_i^{\alpha} Q^{\alpha}\right)_{ij} + V_iq_i^{\alpha}\,,
\end{equation}
with the densities $\rho^{\alpha}$ and a source term for each phase $q^{\alpha}$. The volume of the pore body is denoted by $V_i$.\\
These equations simplify for incompressible fluids with no local sources and sinks to the volume balance for each phase $\alpha$ in a pore body $i$ with the fluxes through the connected throats $Q_{ij}$ to
\begin{equation}
V_i^{\alpha}\frac{\Delta s_{\alpha}}{\Delta t} = \sum_j Q_{ij}^{\alpha}\,,
\end{equation}
with the change in phase saturation per time step $\frac{\Delta s_{\alpha}}{\Delta t}$.\\
Following \cite{weishaupt2021dynamic}, the volumetric flow rate $Q_{ij}$ from pore $i$ to a neighboring pore $j$ via the pore throat $ij$ is approximated using a Hagen-Poiseuille-type flow for each phase $\alpha \in \left\lbrace n,w\right\rbrace $ through the pore throats
\begin{equation}
	Q_{ij}^{\alpha} = k_{ij}^{\alpha}\left(p_i^{\alpha} -p_j^{\alpha} \right)\,.
\end{equation}
Where the flow is proportional to the difference of the phase pressures $p_i^{\alpha}$ and $p_j^{\alpha}$ at the centers of the respective pore bodies (Fig.\ref{fig:PNM-flux}) and a flow conductance factor $k_{ij}^{\alpha}$, which is dependent on the fluid properties and characteristic geometric features of the pore throat, e.g. throat length, inscribed radius and geometric shape \cite{weishaupt2021dynamic}.
\newline
The local capillary pressure is defined as the difference between the non-wetting phase pressure and the wetting phase pressure inside one pore body and can be expressed as a function of the wetting phase saturation for each pore body. In our context of a hydrophobic material in domain I, the wetting phase saturation describes the amount of gas in the pore with respect to the pore size ($s_w = s_{gas} = \frac{V_{gas}}{V_{pore}}$). For simplicity, in this work, we define the local capillary pressure as the difference between the water phase pressure and the gas phase pressure
\begin{equation}
	p_{c,\text{local}}^i = p_i^{liq} - p_i^{gas} = f(s^w)\,. 
\end{equation}
This yields to a positive capillary pressure in the hydrophobic domain I ($p_i^{liq} > p_i^{gas}$) and a negative capillary pressure in the hydrophilic domain II ($p_i^{liq} < p_i^{gas}$).\\
The local capillary pressure saturation relation, $p_{c,\text{local}} (s_w)$, is determined geometrically for cubic pores \cite{joekar2010non}:
\begin{equation}
	p_{c,\text{local}}^i(s_w) = \frac{2\sigma^{nw} \cos \theta}{R_i \left(1-\exp \left(-6.83 s^w \right)\right)}\,,
	\label{eq:pcsw}
\end{equation}
with the surface tension $\sigma^{nw}$, the wetting angle $\theta$ and the pore body inscribed radius $R_i$.
The capillary entry pressure for throats with a square shaped cross-sectional area is derived from the change in free energy for a displacement of the fluid meniscus in a throat following \cite{oren1998extending,blunt2017multiphase}:
\begin{equation}
	p_{c,\text{local}}^{entry} = \frac{\sigma^{nw} \cos \theta \left(1 + 2 \sqrt{\pi G}\right)}{r_{ij}} F_d (\theta, G)
\end{equation}
with the dimensionless function
\begin{equation}
F_d(\theta, G) = \frac{1+\sqrt{1+4GD/\cos^2 \theta}}{1+2\sqrt{\pi G}}\,,
\end{equation}
and
\begin{equation}
G =\frac{A_{ij}}{P_{ij}^2}\,, \qquad
D = \pi -3\theta + 3 \sin \theta \cos \theta -\frac{\cos^2 \theta}{4G}\,.
\end{equation}
The shape factor is denoted as $G$, $A_{ij}$ is the cross-sectional area of the throat between pore $i$ and $j$, and $P_{ij}$ is the corresponding perimeter of that cross-sectional area.\\
Due to the fibrous structure of the material, the void spaces (pores) have an angular structure, where corner flow is relevant. As a geometric simplification, which includes this relevant effect, a cubic approximation is used for the pore bodies and we use throats with a square cross-sectional area.
Following \cite{joekar2010non} and \cite{blunt2017multiphase}, the resulting conductivities of the non-wetting phase in the throats are
\begin{equation}
\label{eq:kn}
	k_{ij}^n = \frac{\pi}{8\mu^nl_{ij}}\left(r_{ij}^\mathrm{eff} \right)^4 \qquad \mathrm{with} \qquad r_{ij}^\mathrm{eff}=\sqrt{\frac{4}{\pi}}r_{ij}\,,
\end{equation}
with the viscosity of the non-wetting phase $\mu^n$ and the length of the throat $l_{ij}$.\\
And for the wetting phase conductivities in the throats, we get
\begin{equation}
	k_{ij}^w = \frac{4-\pi}{\beta 8\mu^wl_{ij}}\left(r_{ij}^{c} \right)^4 \qquad \mathrm{with} \qquad r_{ij}^{c}=\frac{\sigma^{nw}}{p_{ij}^c}\,,
\end{equation}
with the wetting phase viscosity $\mu^w$, the capillary pressure in the upstream pore of the throat $p_{ij}^c$ and the surface tension between the phases $\sigma^{nw}$.\\
The factor $\beta$ is a geometry dependent crevice resistance factor, which is calculated following \cite{Zhou1997} by
\begin{equation}
\beta = \frac{ 12 \sin^2(\gamma)\left(1-B \right)^2}{\left(1-\sin(\gamma)\right)^2 B^2} \cdot 
\frac{\left(\phi_1 - B\phi_2 \right) \left(\phi_3 + B \phi_2 - \left(1-B\right)r_{ij}\right)^2}{\left(\phi_1 - B \phi_2 -\left(1-B \right) r_{ij}^2 \right)^3}
\end{equation}
with $B = \left(\frac{\pi}{2}-\gamma\right)\tan(\gamma)$, $\phi_1 = \cos^2(\gamma + \theta)+ \cos(\gamma + \theta)\sin(\gamma + \theta)\tan(\gamma)$, $\phi_2 = 1-\frac{\theta}{\pi/2 -\gamma}$, and $\phi_3 = \frac{\cos(\gamma + \theta)}{\cos(\gamma)}$. Where $\gamma$ denotes the half corner angle. The formulation is an approximate solution derived using an approach which combines the analytical solutions for the hydraulic diameter and the thin-film flow approximations to reduce the errors of the single approximate analytical solutions \cite{Zhou1997}.\\
\subsection{Interface model concepts}\label{sec:InterfaceConcept}
\begin{figure}[h!]
	\includegraphics[width=0.4\textwidth]{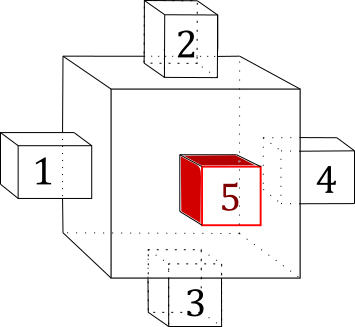}
	\caption{Cutout of the network with the interaction pore which has hydrophobic walls (black) and one hydrophilic throat is connected to that pore (red).}
	\label{fig:InteractionPore}
\end{figure}

\begin{figure}[h!]
	\includegraphics[width=0.7\textwidth]{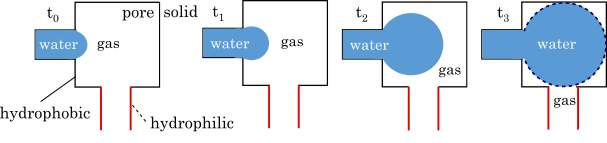}
	\caption{Fluid configurations of water invasion in a hydrophobic boundary pore connected to a hydrophilic throat (red) at different time steps $t_0 < t_1 < t_2 <t_3$ during the invasion process. The dashed state at $t_3$ marks the time, when the water gets in touch with the hydrophilic wall.}
	\label{fig:Config1-timesteps}
\end{figure}

\begin{figure}[h!]
	\includegraphics[width=0.5\textwidth]{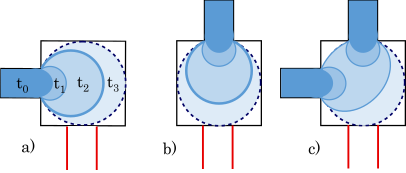}
	\caption{Different configurations of water invasion in a hydrophobic boundary pore connected to a hydrophilic throat (red).}
	\label{fig:Config1-geometries}
\end{figure}

\begin{figure}[h!]
	\includegraphics[width=0.8\textwidth]{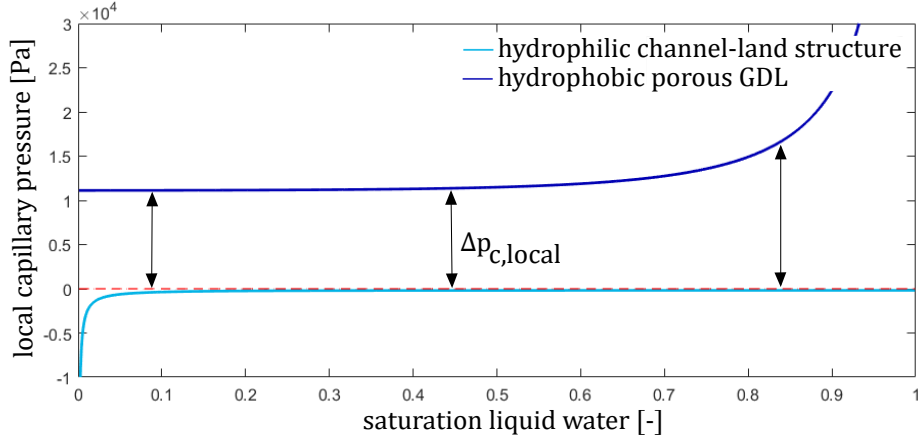}
	\caption{Local pore-scale capillary pressure saturation relation for a hydrophilic (light blue) (domain II) and a hydrophobic (dark blue) domain (domain I). The capillary pressure is defined as the difference between the non-wetting phase pressure (here: water) and the wetting phase pressure (here: gas). The water saturation describes the amount of water in the pore with respect to the pore size ($s_{liq} = \frac{V_\mathrm{liquid}}{V_\mathrm{pore}}$). In the GDL, the water saturation equals the non-wetting phase saturation $s_n$, while in the gas distributor channels, the water saturation is the wetting phase saturation $s_w$.}
	\label{fig:pCS-curves}
\end{figure}

\begin{figure}[h!]
 	\includegraphics[width=0.5\textwidth]{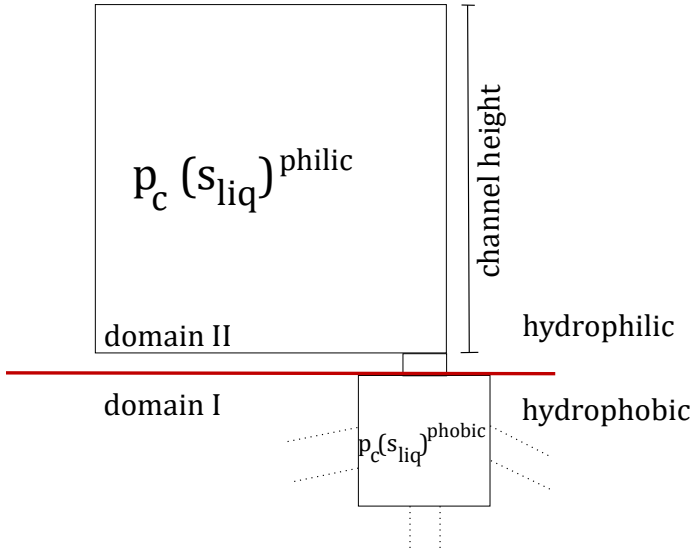}
 	\caption{Representation of two pores at the interface between the hydrophobic (domain I) and the hydrophilic domain (domain II). The hydrophilic channel is represented by a cubic pore with equal side dimensions to define a local capillary pressure saturation relation. In each of the pores a different capillary pressure saturation relation with respect to the water saturation $s_{liq}$ is used.}
 	\label{fig:PcSwInterface}
 \end{figure}

\begin{figure}[h!]
	\includegraphics[width=0.5\textwidth]{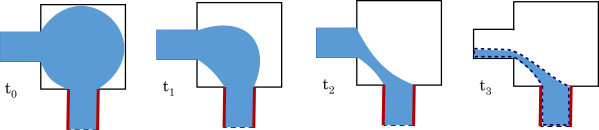}
	\caption{Different configurations of water (non-wetting phase in the pore) leaving the hydrophobic pore through the hydrophilic throat (red) at different time steps during the emptying process ($t_0 <t_1<t_2<t_3$).}
	\label{fig:Config2-timesteps}
\end{figure}

\begin{figure}[h!]
	\includegraphics[width=0.5\textwidth]{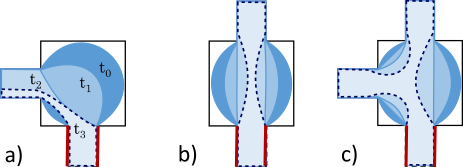}
	\caption{Different configurations of water (non-wetting phase in the pore) leaving the hydrophobic pore through the hydrophilic throat (red) for different geometric configurations. The color shades refer to the different time steps (from dark to light).}
	\label{fig:Config2-geometries}
\end{figure}

We consider the interaction of fluids in a hydrophobic pore network (domain I) interacting with a hydrophilic domain (domain II) representing a hydrophobic GDL and a hydrophilic gas distributor of a PEM fuel cell, respectively. It is assumed that the flow in the hydrophobic GDL (domain I) is not influenced by the hydrophilic gas distributor (domain II) until water gets in touch with hydrophilic surface in the interface pores. For the interface, we assume a pore body with hydrophobic walls. A cutout of the pore network with the interaction pore is shown in Fig.\ref{fig:InteractionPore}.
\newline
One of the throats connected to that pore is assumed to have hydrophilic walls (red). The water displaces the gas phase in the pore body coming from the connected hydrophobic throats (throats 1-4). The occuring water-gas configurations depend on the geometry and wetting behavior of this pore. In Fig.\ref{fig:Config1-timesteps}, the fluid configurations during the invasion process at different time steps in a basic example geometry are visualized in two dimensions.
\newline
The invading water forms a growing sphere inside the cubic pore body. The gas phase is displaced by the water. At a certain time, the water gets in touch with the hydrophilic wall. The moment immediately before the contact is marked with a dashed line in Fig.\ref{fig:Config1-timesteps}. 
We only consider cubic pores connected at their face centers by throats with a square cross sectional area. In Fig.\ref{fig:Config1-geometries}, it is shown that the water configuration immediately before the contact with the hydrophilic throat is independent of the connections and geometric configuration of the cubic pore.
\newline
The different configurations show the water filling of a hydrophobic pore (black line) connected to a hydrophilic throat (red line). The dashed state marks the time, immediately before the water gets in touch with a hydrophilic wall and the water behavior is influenced by the hydrophilic domain II. The different shades mark the fluid configurations at different time steps ($t_0 < t_1 < t_2 <t_3$). It can be seen that in the configurations a), b) and c), the non-wetting water touches the hydrophilic throat approximately at the same filling state even though the connections are applied differently. Configuration c) shows that even several connecting (and filling) throats do not influence the water configuration. This results from the assumption that for strongly non-wetting pores, the water forms a sphere in the pore body \cite{joekar2010non}. The cubic pore needs to be filled by a complete sphere before the water "sees" the hydrophilic throat. For cubic pores the water saturation is $s_{liq} = \frac{V_{sphere}(R_i)}{V_{pore}} = 0.52$ at this state.\\
The independence of the filling state from the connectivity configuration allows to take the local water saturation $s_{liq}$ as a threshold criterion at the interface pores (at the interface between the hydrophobic porous domain I and the hydrophilic domain II).\\
After reaching the hydrophilic throat, the fluid distribution will be influenced by the connected hydrophilic domain II. For both domains (I and II), a local, pore-scale capillary pressure saturation relation $p_{c,\text{local}}^i(s)$ can be formulated based on the pore sizes, pore shapes and contact angles. In this work, the formulation presented by Joekar-Niasar et al. \cite{joekar2010non} for cubic pores is used (Eq.\ref{eq:pcsw}). In Fig.\ref{fig:pCS-curves}, the $p_{c,\text{local}}(s)$ curves are visualized exemplary for a hydrophobic domain I such as the GDL and a hydrophilic domain II such as a gas distributor in PEM fuel cells. Here, the channels of the gas distributors are considered as large pores with a diameter of 1mm corresponding roughly to the width and height of the channels. This yields to a similar curvatures of the fluid interface in this pore as in the gas distrirbutor channel and the resulting capillary pressure-saturation relation is comparable.
\newline
 Note that in Fig.\ref{fig:pCS-curves}, the local capillary pressure is plotted with respect to the water saturation, which is the wetting phase saturation in the hydrophilic channel-land structure (domain II) and the non-wetting phase saturation in the GDL (domain I).
 Dependent on the water saturation in the boundary pore, a capillary pressure difference ($\Delta p_{c,\text{local}}$ in Fig.\ref{fig:pCS-curves}) results between the hydrophobic (domain I) and the hydrophilic domain (domain II). To formulate the local capillary pressure saturation relation in the hydrophilic channel, a cubic pore with the same side length as the channel cross-section is considered (Fig.\ref{fig:PcSwInterface}). The geometry of this representative pore 293 gives a comparable phase interface curvature with the original geometry, and therefore, its capillary pressure saturation relation is used representing the pressure configuration in the channel (domain II). The resulting capillary pressure saturation relation is used representing the pressure configuration in the channel (domain II).
 Since the representing pore of the hydrophilic channel (domain II) are much larger than the hydrophobic pores in the GDL (domain I) ($R_{channel} = 0.5mm, R_{GDL} < 100\mu m$), the local capillary pressure curve is steeper and reaches a value very close to zero even for small water saturations $s_{liq}$. Therefore, as a model assumption, a constant capillary pressure in the hydrophilic domain II is applied. This simplification is based on the assumption that the water is transported to the wall of the channel opposite to the porous GDL (see Fig.\ref{fig:CathodeConfigurations}) and away from the interface between the domains due to the hydrophilic wall properties in the gas channels. Nevertheless, for a more general formulation, a water saturation dependent capillary pressure $p_{c,\text{local}}(s_{liq})$ can be implemented for the hydrophilic domain II. This might be necessary, if very high water saturations $s_{liq}$ are considered in the channels, which might occur during local blocking of the gas channels. However, this situations should be avoided at the operating conditions of the fuel cell such that it is not in the scope of this work.\\
As previously mentioned, the water saturation $s_{liq}$ in the pore needs to reach a certain value to start an interaction between the hydrophobic domain I and the hydrophilic domain II.\\
The capillary pressure difference $\Delta p_{c,\text{local}}=p_{c,\text{local}}(s_{liq})^{phobic}-p_{c,\text{local}}(s_{liq})^{philic}$ (Fig.\ref{fig:PcSwInterface}) between the hydrophobic GDL pores (domain I) and the hydrophilic gas channel (domain II) shows the significance of applying a saturation threshold for the interaction. Without the threshold, due to the water pressure gradient, every time step the total amount of water present in the interface pore would flow into the hydrophilic domain II. This would not represent a physical behavior.\\
Once the water saturation (here: non-wetting phase) threshold $s_{liq}^{th}$ is fulfilled (contact of water with the hydrophilic throat), the water flows out of the pore (in domain I). The flux is defined by the water pressure difference (non-wetting phase in GDL) between the considered pore (in domain I) and an assumed water pressure in the connected hydrophilic channel-land structure (domain II) (water is wetting phase in gas distributor). As shown in Fig.\ref{fig:Config1-geometries}, for cubic pores, this saturation threshold is $s_{liq} = \frac{V_{sphere}(R_i)}{V_{pore}} = 0.52$. \\
Additionally to the starting threshold, a stop criterion needs to be formulated. Again, different configurations during the emptying process are considered (Fig.\ref{fig:Config2-timesteps}).
\newline
The dashed state marks the time before the phase is separated and no continuous water phase exists, which connects the hydrophobic network (domain I) with the hydrophilic gas channels (domain II).\\
In contrast to the filling process (Fig.\ref{fig:Config1-geometries}), the emptying process is not independent of the geometric configuration and connections of the pore as shown in Fig.\ref{fig:Config2-geometries}.
\newline
For the different configurations we see very varying fluid patterns. During the emptying process, the water trys to stay connected. The pore body is "emptied" while the water sticks in the hydrophilic pore throat such that it stays water-filled. The hydrophobic throats connected to the considered pore are also influenced by the outflow through the hydrophilic throat. We assume that the hydrophobic throats are emptied as well and the meniscus moves to its other end connected to the next pore body, which stays uninfluenced.\\
Compared to the starting criterion, it is more difficult to formulate a criterion, when a snap-off is happening in the interaction pore such that the water flux is stopped due to a disconnection of the water phase to the hydrophilic domain II. For the beginning, we choose a water saturation threshold $s_{liq}^{th}$ (non-wetting phase) analogous to the start criterion. The flux of the water stops once the water saturation $s_{liq}$ (non-wetting phase saturation) in the considered pore (in domain I) is lower than this certain value. For cubic pores, we choose $s_{liq}^{stop} = 0.005$. In the applications, we will see that the starting criterion has a dominant influence on the behavior, while the exact value of the stopping criterion is less relevant.\\ \\
Using the theoretical considerations presented above, we formulate a combined criterion of pressure and water saturation $s_{liq}$ (non-wetting phase in hydrophobic porous domain I) configuration at the interface between the hydrophobic porous domain I and the hydrophilic domain II, which defines the interaction and the influence on the flow behavior:
\begin{itemize}
	\item the flow is only influenced once the water configuration in the boundary pore "touches" the hydrophilic throat, 
	\item once the throat is reached, flow is defined by the water pressure differences between the hydrophobic porous domain I and the hydrophilic domain II,
	\item the flow is stopped once the connection between the hydrophilic throat and the "bulk" water is lost (snap-off occurred in the pore body).
\end{itemize}
Here, the gas phase pressure is nearly constant such that the water phase pressure difference equals the capillary pressure difference. Therefore, the flux across the interface between the hydrophobic porous domain I and the hydrophilic domain II, through the hydrophilic throat, is driven by the capillary pressure difference $\Delta p_{c,\text{local}}$ across the interface (Fig.\ref{fig:pCS-curves}).
\section{Numerical model}
In the pore-network model, a volume balance equation for each phase $\alpha \in \left\lbrace n,w\right\rbrace$ for each pore body $i$ is solved
\begin{equation}
	\sum_{j=1}^{N_i} q_{ij}^\alpha = 0\,.
\end{equation}
$N_i$ is the number of throats connected to pore body $i$.
The volumetric fluxes $q_{ij}$ through throat $ij$ are calculated by means of the Washburn equation
\begin{equation}
	q_{ij} = k_{ij}^\alpha (p_i^\alpha - p_j^\alpha)\,,
\end{equation}
with the throat conductivities $k_{ij}^\alpha$ and the phase pressure difference between the connected pore bodies $p_i^\alpha - p_j^\alpha$.
The balance equations are formulated per pore body, while the fluxes occur within the one-dimensional pore throats \cite{weishaupt2021dynamic}. The primary variables of the pore-network model live on the network nodes, i.e. the center of the pore bodies.\\
The pore bodies located at the interface (interface pores) are used to impose the coupling conditions for the pore-network model by connecting ghost nodes representing the hydrophilic domain II. A Neumann condition is applied which depends on the fixed pressure in the ghost node $p_0$ and in the interface pore in the hydrophobic domain I $p_i$
\begin{equation}\label{eq:qOut}
	q_{out}=-k_{int} \left( p_i^n -p_0 \right) \,.
\end{equation}
Similar to the general pore-network model (Eq.\ref{eq:kn}), the hydrophilic throat has a hydraulic resistance based on its geometric properties:
\begin{equation}
	k_\mathrm{int} = \frac{\pi}{8\mu^{liq}l_\mathrm{int}}\left(r_\mathrm{int}^\mathrm{eff} \right)^4 \qquad \mathrm{with} \qquad r_\mathrm{int}^\mathrm{eff}=\sqrt{\frac{4}{\pi}}r_\mathrm{int}\,,
\end{equation}
with the inscribed radius and the length of the interaction throat, $r_{int}$ and $l_{int}$, respectively. Here, $\mu^{liq}$ denotes the viscosity of the water phase which is the wetting phase in the interface throat (red in Figures \ref{fig:Config1-timesteps}-\ref{fig:Config2-geometries}) for the considered setup describing the interaction of a hydrophobic GDL (domain I) with a hydrophilic gas distributor (channel-land structure) (domain II).\\
The model is implemented in the DuMu$^x$ framework \cite{flemisch2011dumux,koch2020dumux} and embedded in the pore-network model developed by Weishaupt \cite{WeishauptDiss}. The details of the pore-network model implementation can be found in \cite{WeishauptDiss}.
\subsection{Implementation}
The calculation is done using Newton iterations solving the non-linear equations for each time step. An adaptive timestepping is used where the decision if a throat is invaded by the non-wetting phase or not is made during the Newton steps and the time step size is controlled to achieve convergence. For detailed information on the implementation, it is referred to Chen et al., who describe an analogue algorithm to solve dynamic pore-newtork models in \cite{chen2020fully}.
\section{Numerical studies}
\subsection{Fundamental behavior of the interacting pores}
\begin{figure}[h!]
	\includegraphics[width=\textwidth]{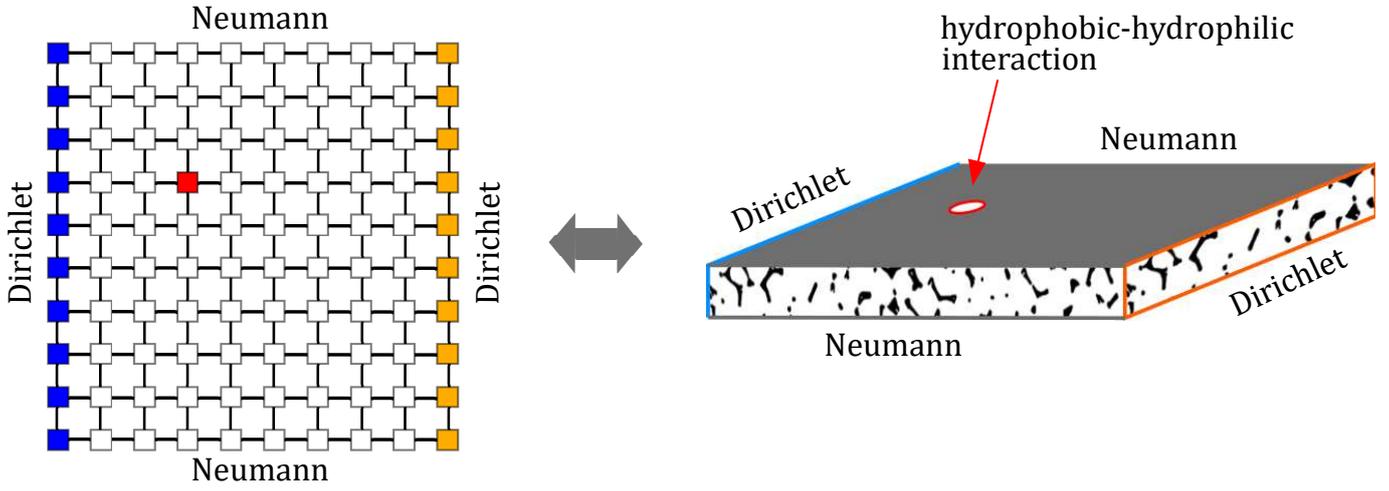}
	\caption{Visualization of the investigated hydrophobic network (left) which is an exemplary representation of the porous medium setup on the right (domain I). The basic setup is built to show the fundamental behavior of the interacting pores. The hydrophobic porous medium (right) is placed between hydrophobic coated plates (the material of the plates itself is hydrophilic) with a hole at the position of the interaction pore. The inner surface of the hole is hydrophilic (red). The outer surface of the plate is hydrophilic too (domain II). The water flow goes from left (blue inlet) to right (orange outlet) (right).}
	\label{fig:10x10-1Pore-Labels}
\end{figure}

\begin{figure}[h!]
	\includegraphics[width=0.9\textwidth]{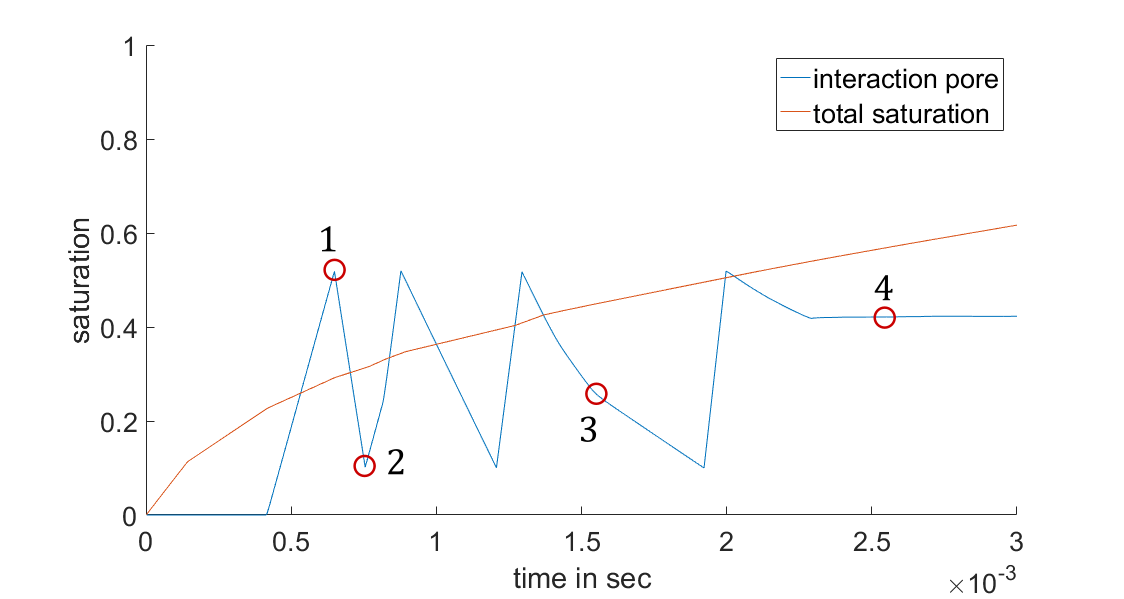}
	\caption{Saturation $s_{liq}$ in the interface pore (blue) and total water saturation $S_{liq}$ of the network (red) over time during the interaction with the hydrophilic domain II. The four states marked red are investigated in detail in the following.}
	\label{fig:10x10-1Pore-stop001-homogeneous-saturation_hydrophilicBC}
\end{figure}

\begin{figure}[h!]
	\includegraphics[width=0.8\textwidth]{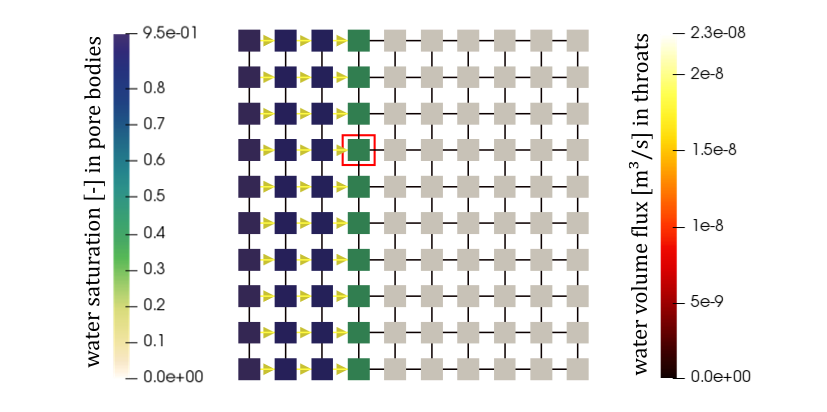}
	\caption{State 1: Water flux in throats and water saturation in pore bodies at $t=0.0006s$. The interaction pore is marked with a red square.}
	\label{fig:1Pore-State1}
\end{figure}

\begin{figure}[h!]
	\includegraphics[width=0.8\textwidth]{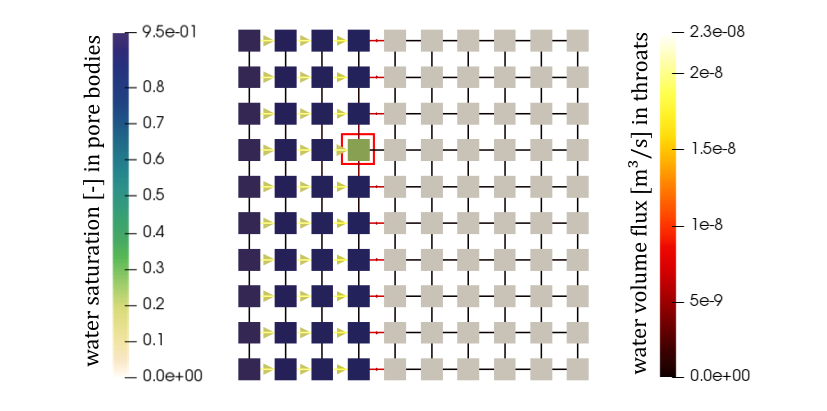}
	\caption{State 2: Water flux in throats and water saturation in pore bodies at $t=0.000815s$. The interaction pore is marked with a red square.}
	\label{fig:1Pore-State2}
\end{figure}

\begin{figure}[h!]
	\includegraphics[width=0.8\textwidth]{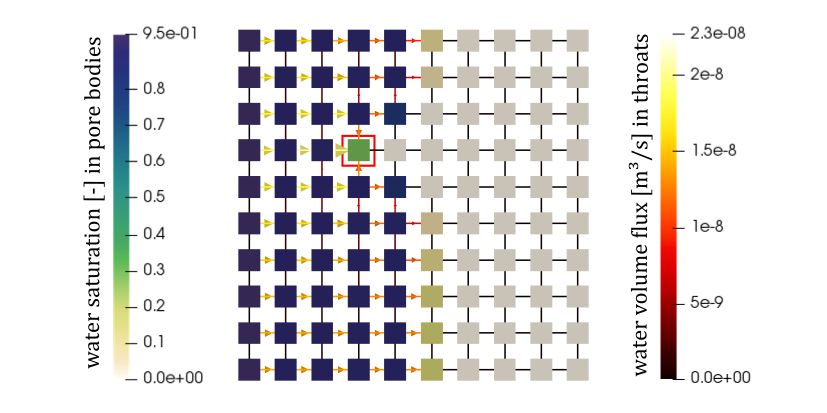}
	\caption{State 3: Water flux in throats and water saturation in pore bodies at $t=0.0015s$. The interaction pore is marked with a red square.}
	\label{fig:1Pore-State3}
\end{figure}

\begin{figure}[h!]
	\includegraphics[width=\textwidth]{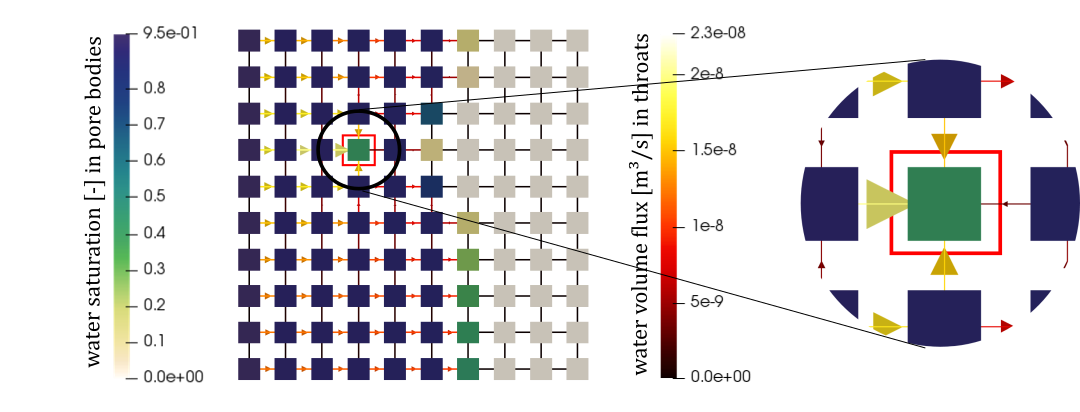}
	\caption{State 4: Water flux in throats and water saturation in pore bodies at $t=0.0025s$. The interaction pore is marked with a red square.}
	\label{fig:1Pore-State4}
\end{figure}

To investigate the fundamental behavior of the system, a basic, homogeneous, two-dimensional, hydrophobic network (domain I) is considered (Fig.\ref{fig:10x10-1Pore-Labels} (left)). Imagine a thin porous medium between two hydrophobic coated plates. In one of the plates, a hole is drilled. The inner surface of the drilling hole is now uncoated and therefore hydrophilic. The position of the drilling hole denotes the position of the interaction pore (Fig.\ref{fig:10x10-1Pore-Labels} (right)). The outer surface of the plate is hydrophilic (domain II). In the representing pore network (domain I), the pore bodies and throats are of uniform size (pore size ($R_i = 10^{-4}$m), throat size ($r_{ij} = 0.5\cdot 10^{-4}$m) and length ($l_{ij} = 3.3\cdot 10^{-3}$m)). The domain is initially filled with the gas phase ($S_{gas} = 1.0$) and a primary drainage process of the water phase displacing the gas phase in the entire domain is considered in the hydrophobic domain I. The blue pores denote the inlet pores, where a water saturation $s_n = s_{liq}= s_{inlet}$ is fixed as a Dirichlet boundary condition. The orange pores are the outlet pores, where the gas saturation is fixed to one $s_{gas} = s_w = 1.0$. No pressure difference is applied for the gas (wetting phase) pressure ($\Delta p^w = p^{gas}= 0$). The red pore is connected to the hydrophilic domain II. Here, an outflux is applied and suppressed dependent on the fulfilled water saturation thresholds. The applied time step size is $\Delta t = 10^{-5}$s. 
\newline
In Fig.\ref{fig:10x10-1Pore-stop001-homogeneous-saturation_hydrophilicBC}, the water saturation (non-wetting phase) in the pore (in domain I) interacting with the hydrophilic domain II (interaction pore) (blue line) and the global saturation of the network (red line) during the invasion process are shown. The interface pore (in domain I) is filled and emptied with water due to the interaction with the hydrophilic domain II based on the applied water (non-wetting phase) saturation thresholds ($s_{n,liq}^{start} = 0.52$ and $s_{n,liq}^{stop} = 0.005$). The total water saturation $S_{liq}$ increases monotonously, while the saturation in the interaction pore shows a varying, oscillating behavior. The volume flux leaving the hydrophobic domain I is very small compared to the overall water flux through the system. Therefore, the outflux through the interaction pore has only a small influence on the overall (global) saturation of the network.
In the following, the four marked states in Fig.\ref{fig:10x10-1Pore-stop001-homogeneous-saturation_hydrophilicBC} are described and explained in detail to understand the behavior of the water (non-wetting phase) saturation curve corresponding to the interaction pore (in domain I).\\
\newline
Figures \ref{fig:1Pore-State1}-\ref{fig:1Pore-State4} show the volume flux of the water through the connecting throats in the hydrophobic network (domain I).\\
At state 1 (Fig.\ref{fig:1Pore-State1}), the water saturation (non-wetting phase) threshold ($s_{n/liq} = 0.52$) in the pore (in domain I) connected to the hydrophilic domain II is reached and the interaction between the hydrophobic network (domain I) and the hydrophilic domain (domain II) starts through this pore such that the pore-local water phase saturation $s_{liq}$ decreases.\\
\newline
At state 2, the water saturation (non-wetting phase) in the pore (in domain I) connected to the hydrophilic domain II falls below the threshold of the stop criterion ($s_{n/liq} = 0.005$) and the flux between the hydrophobic pore network (domain I) and the hydrophilic domain (domain II) stops (State 2, Fig.\ref{fig:1Pore-State2}).\\
\newline  
This small saturation leads to a fast refilling of the interaction pore. This occurs again before state 3. At state 3, the decrease of the water saturation (non-wetting phase) in the pore (in domain I) connected to the hydrophilic domain II is halted due to the flux from neighboring pores (throats 1,2 and 3 in Fig.\ref{fig:InteractionPore}, and denoted by arrows in Fig.\ref{fig:1Pore-State3}), where the capillary entry pressure, which is necessary to overcome, to invade the connecting throat is met (State 3, Fig.\ref{fig:1Pore-State3}). \\
\newline
Once the downstream neighboring throat (throat 4 in Fig.\ref{fig:InteractionPore}) is invaded, water flows through this throat as well into the interaction pore following the local water pressure gradient. This results in a filling from all four hydrophobic throats connected to the interaction pore. In the considered test case, this results in an equilibration of the fluxes in and out of the interaction pore and the saturation stays constant (State 4, Fig.\ref{fig:1Pore-State4}).\\
\subsection{Investigation of the network response to interaction of several interface pores}
\begin{figure}[h!]
	\includegraphics[width=0.9\textwidth]{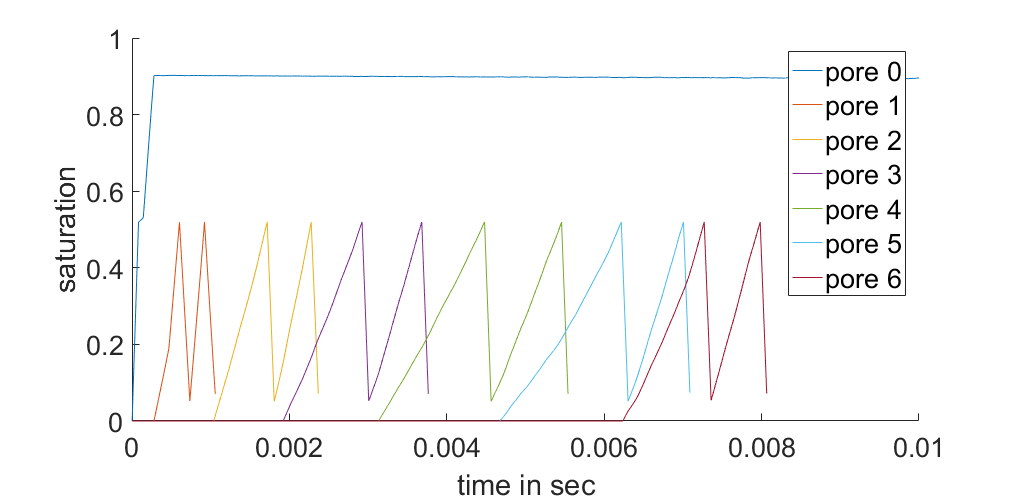}
	\caption{Water saturation in the interface pores at the top face during the interaction with the hydrophilic domain II.}
	\label{fig:TopPores-saturations}
\end{figure}

\begin{figure}[h!]
	\includegraphics[width=0.8\textwidth]{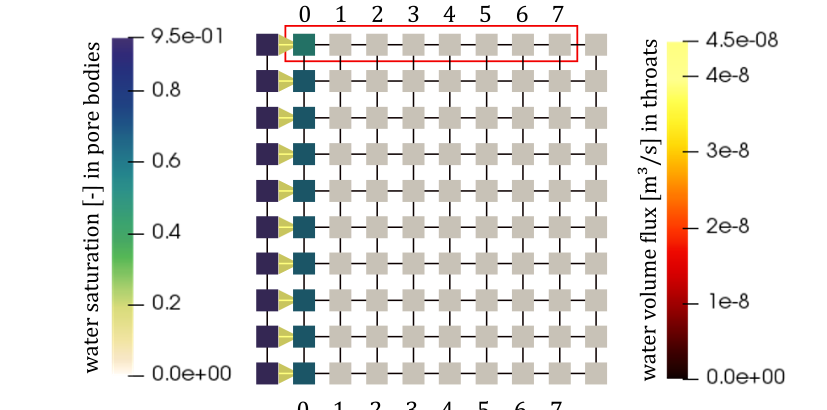}
	\caption{Water volume flux in $[m^3/s]$ in throats and water saturation $[]$ in pore bodies at $t=0.0001s$)}
	\label{fig:TopPores-State1}
\end{figure}

\begin{figure}[h!]
	\includegraphics[width=0.8\textwidth]{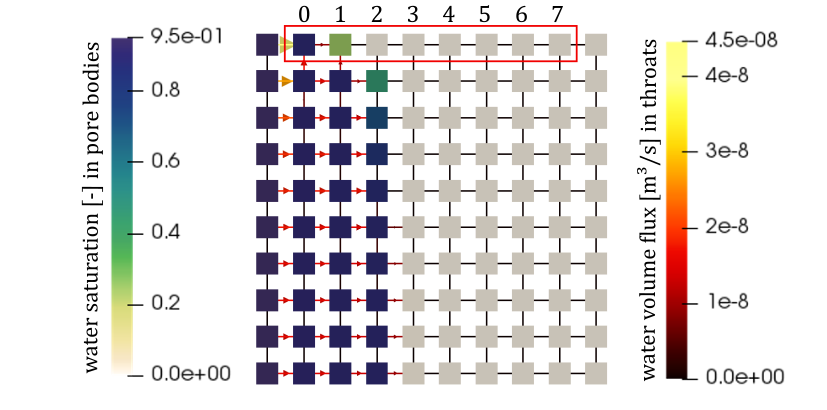}
	\caption{Water volume flux in $[m^3/s]$ in throats and water saturation $[]$ in pore bodies at $t=0.00082s$)}
	\label{fig:TopPores-State2}
\end{figure}

\begin{figure}[h!]
	\includegraphics[width=0.9\textwidth]{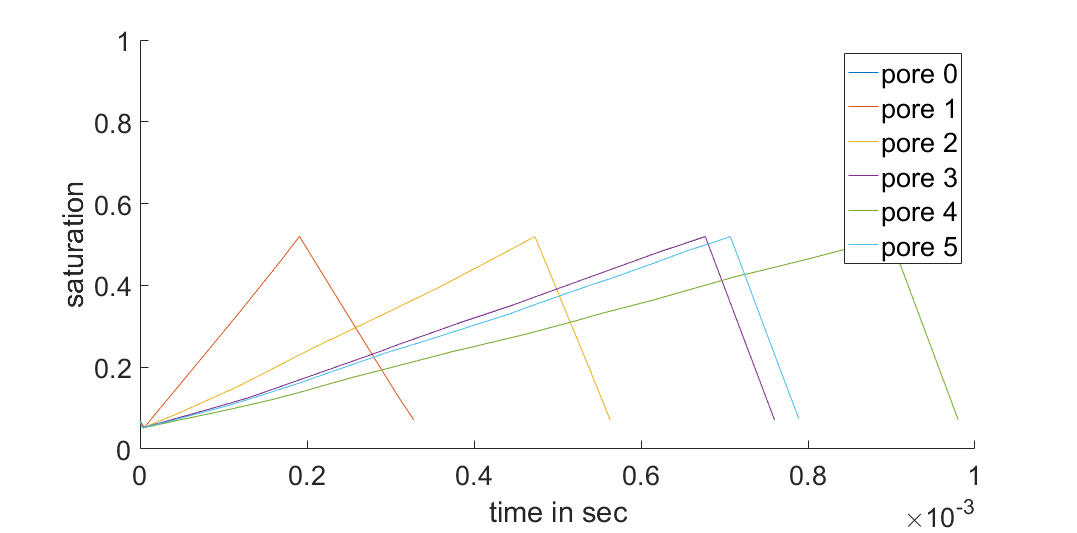}
	\caption{The time dependent water saturation $s_{liq}$ evolution shows the different filling velocity of the interaction pores. For the first five pores the second oscillation is shown to compare the frequency of the water saturation oscillations in the pores. Here, only the first water filling of each pore is plotted. Pore 0 is missing since the water saturation does not oscillate but reaches an equilibrium (see Fig.\ref{fig:TopPores-saturations})}
	\label{fig:TopPores-compareFreq}
\end{figure}

\begin{figure}[h!]
	\includegraphics[width=0.8\textwidth]{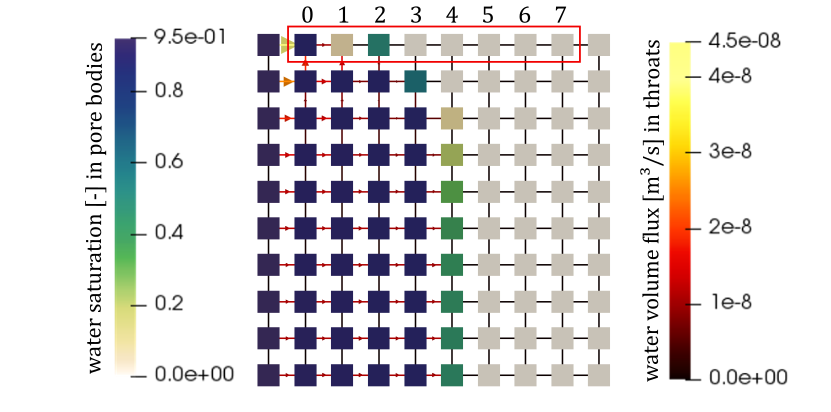}
	\caption{Water volume flux in $[m^3/s]$ in throats and water saturation $[]$ in pore bodies at $t=0.0017s$)}
	\label{fig:TopPores-State3}
\end{figure}

\begin{figure}[h!]
	\includegraphics[width=0.8\textwidth]{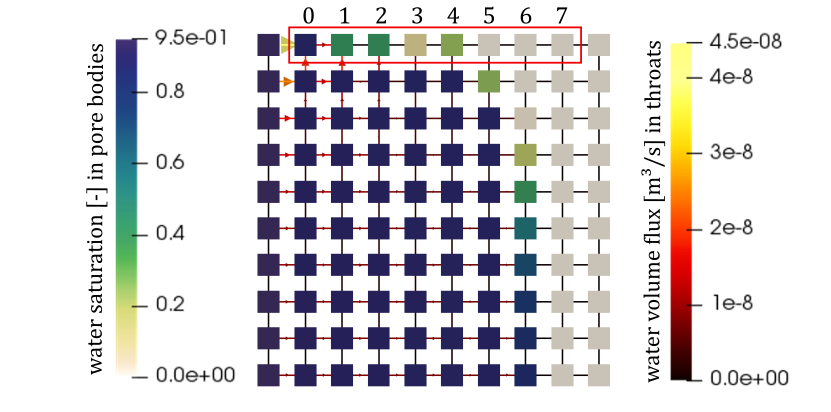}
	\caption{Water volume flux in $[m^3/s]$ in throats and water saturation $[]$ in pore bodies at $t=0.0038s$)}
	\label{fig:TopPores-State4}
\end{figure}

The presented model does not only capture the interaction of one single pore (in domain I) with hydrophilic domain II but allows an arbitrary combination of interface pores. In the presented example, the complete pore row at the top of the network (domain I) is in contact with hydrophilic domain II.\\
The interface pores are located at the top of domain I and are in contact with an gas-filled (no water present) hydrophilic domain II. The interface pores are framed red and numbered consecutively in Figures \ref{fig:TopPores-State1}-\ref{fig:TopPores-State4}. It is assumed that the water distributes fast in the hydrophilic domain II and is transported away from the interface between the hydrophobic porous domain I and the hydrophilic domain II such that there is no influence on the outflow behavior between the different interface pores.\\
As in the previous example, water is used as the non-wetting fluid and pushed through the domain from left to right (Fig.\ref{fig:10x10-1Pore-Labels}). The hydrophobic domain I is initially filled with air (gas). In Fig.\ref{fig:TopPores-saturations}, the oscillating water saturations in the interface pores are shown. For transparency reasons, only the first two oscillations are plotted.
\newline
The water saturation in the first interface pore in the top row (interface pore 0) does not show an oscillation. We see a kink in the curve at a water saturation of $s_{liq} = 0.52$, which is the threshold for the interaction with the hydrophilic domain II. At this time, an outflux through this pore from the hydrophobic domain I to the hydrophilic domain II starts. The fluid configuration in the network (domain I) during this time step is shown in Fig.\ref{fig:TopPores-State1}. However, the fluxes into the interface pore from the neighboring pores in the hydrophobic network (domain I) is still higher than the outflux, such that the water saturation keeps increasing. This is a boundary effect which results from the position of the interface near the inlet pores. Pore 0 gets virtually unlimited fresh water from the left edge and an equilibrated water saturation is reached based on the equilibrium of in- and outflux of the interface pore. With increasing water saturation in the pore, the local capillary pressure and water phase pressure rise. This might cause a steeper gradient between the interface pore and the surrounding pores in the hydrophobic network (domain I) but also a larger pressure difference between the hydrophobic domain I and the hydrophilic domain II influencing the fluxes through the hydrophobic and hydrophilic throats. These fluxes equilibrate for interface pore 0 at a constant water saturation.
\newline
The replacement process continues due to the water (non-wetting phase) pressure gradient between the left and right side of the network (domain I) such that the next column of pores is invaded by the non-wetting phase (water). The second interface pore (interface pore 1) shows an oscillating behavior of the water saturation (see Fig.\ref{fig:TopPores-saturations}). In this case, no equilibrium is reached between the in- and outfluxes. The outflow starts with the start threshold at $s_{liq} = 0.52$ and causes an emptying of the pore body. Once the stop threshold is reached, the interaction with the hydrophilic domain II stops and the water saturation increases due to the fluxes through the connecting throats of the hydrophobic network (domain I). The oscillating behavior shows that the outflux from the hydrophobic pore into the hydrophilic domain II is larger than the "refilling" fluxes coming from the surrounding hydrophobic pores (domain I). In Fig.\ref{fig:TopPores-State2}, the time step is captured when interface pore 1 is filled with the non-wetting water for the first time.
\newline
The water fluxes in the hydrophobic throats (in domain I) show that the water does not invade the network straight from left to right as it would be the case in a network without interface pores. We see a clear upwards flux perpendicular to the applied pressure gradient due to the outflux from the interface pores. The resulting decrease of the water saturation causes a gradient of the local capillary pressure such that the interface pores are refilled from the hydrophobic adjacent pores (in domain I).\\
In Fig.\ref{fig:TopPores-State3}, we see how the second interface pore (pore 1) is emptied compared to the previous state (Fig.\ref{fig:TopPores-State2}). At the same time, the third interface pore (pore 2) is invaded. From the water saturations in the top row pores (interface pores), we see that the water saturations are not syncronic oscillating. The time dependent evolution of the water saturations (Fig.\ref{fig:TopPores-compareFreq}) shows that the frequency of the oscillations depend on the position of the interface pore in the network (domain I). Early invaded pores (pores further left), where the pressure gradient is higher during primary drainage, oscillate with a higher frequency. In pore 5, we see the influence of the small size of the network (Fig.\ref{fig:TopPores-compareFreq}). Once the overall network reaches the maximum water saturation, the flux into the interaction pores is increased. The water does not need to be distributed elsewhere. 
\newline
The row of interface pores does not only influence the adjacent pores but the fluid pattern in the complete domain I. In Fig.\ref{fig:TopPores-State4}, we see the deformation of the invasion front resulting from the outflux in the interface pores.
\newline
However, it has to be noticed that this is just a basic example with a small network. We aim to show the complex interplay of the hydrophobic network (domain I) interacting with a non-resolved hydrophilic domain (domain II). The example shows that, with this quite simple approach, the complex interactions can be captured on the pore-scale.
\subsection{Application to the interface between GDL and gas distributor}
\begin{figure}[h!]
	\includegraphics[width=0.5\textwidth]{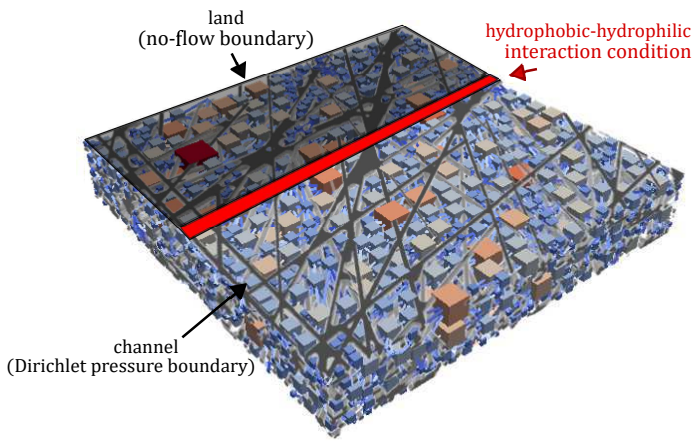}
	\caption{Part of a pore network with cubic pore bodies embedded in the GDL structure, from which the network has been extracted using PoreSpy \cite{gostick2017versatile}. The complete network includes 6976 pores and 31774 throats. The grey area respresents the part where a no-flow boundary condition is applied representing the land part. The red part is the line, where the interaction condition between the hydrophobic GDL (domain I) and the hydrophilic channel-land structure (domain II) is applied. The open area represents the channel. Here, a Dirichlet pressure boundary condition is applied to allow outflow of the phases.}
	\label{fig:GDL-PoreNetwork}
\end{figure}

\begin{figure}[h!]
	\includegraphics[width=0.9 \textwidth]{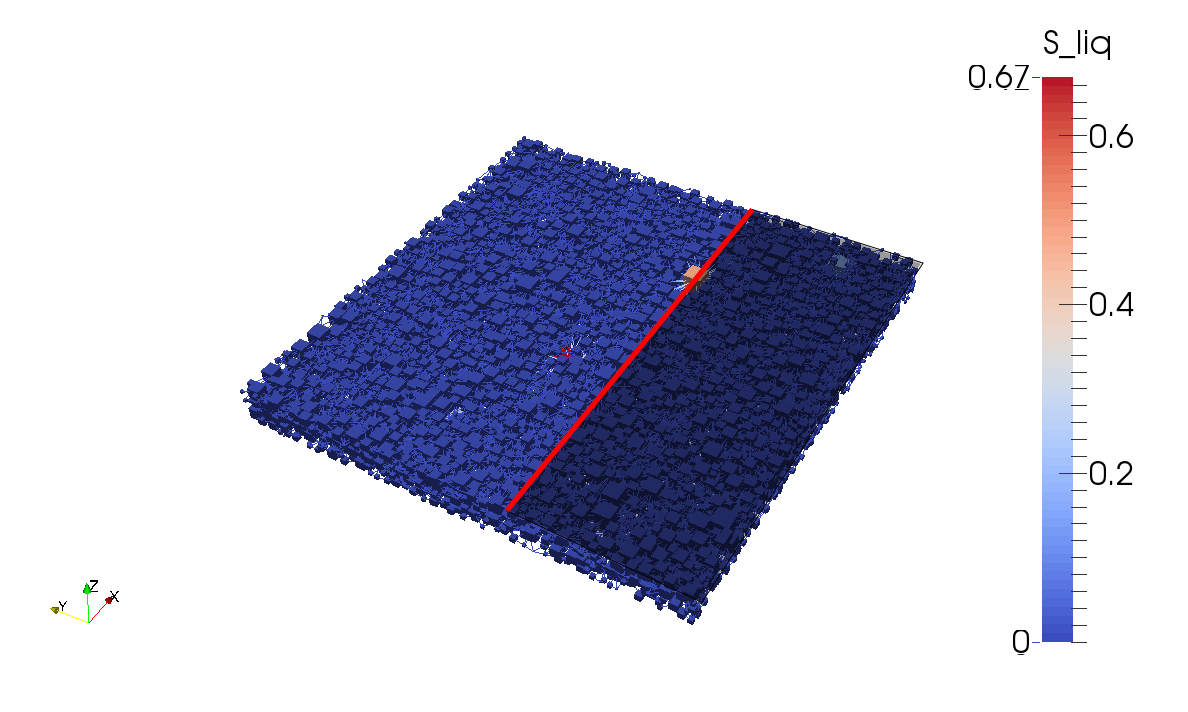}
	\caption{Water saturation (non-wetting phase) in the pores before the breakthrough (saturation threshold not fulfilled).}
	\label{fig:GDL-breakthrough}
\end{figure}

\begin{figure}[h!]
	\includegraphics[width=0.9 \textwidth]{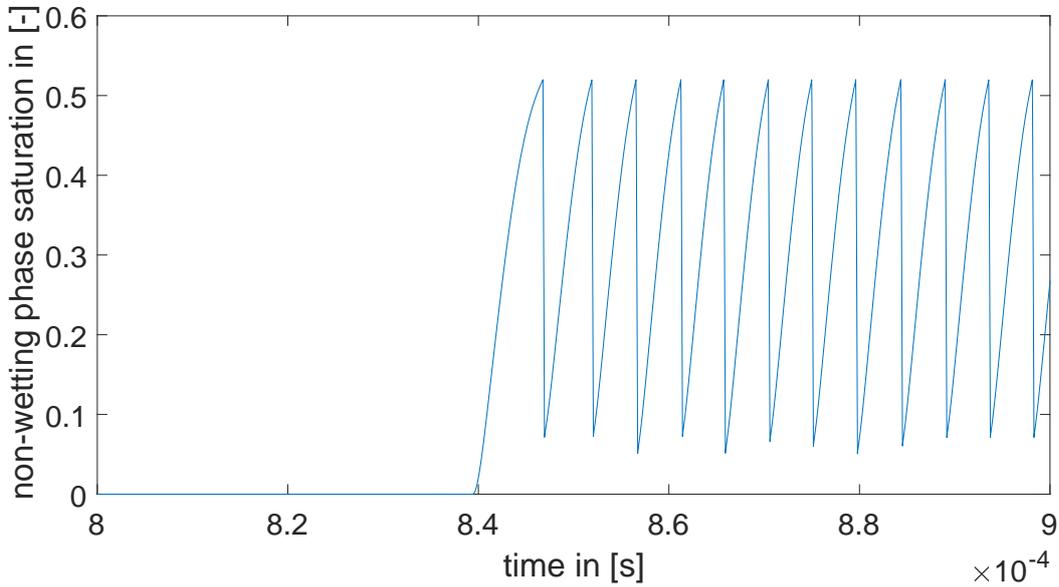}
	\caption{Water saturation in the pore interacting with the hydrophilic domain.}
	\label{fig:GDL-saturation}
\end{figure}

With the developed concept, we want to capture parts of the interaction between the fluid displacements in the GDL (domain I)and the gas distributor flow (domain II). Since the GDL consists of a hydrophobic material and is partly in contact with the hydrophilic gas distributor, local effects of this interaction need to be captured by the model to simulate the flow through the GDL structure (domain I). As an example application, here, the primary displacement of air by water through a pore network extracted from a fibrous GDL representation structure is shown. For the representation of the GDL on the pore-scale, a network of cubic pores is used. The size and connectivity of the pores is based on the void space of the fiber structure. For the network extraction the python-based open-source tool PoreSpy \cite{gostick2017versatile} is used. In Fig.\ref{fig:GDL-PoreNetwork}, a part of the resulting network embedded in the base structure is shown.
\newline
The network (domain I) is initially filled with gas. We are interested in the interaction of the flow in the hydrophobic GDL (domain I) with the hydrophilic wall of the gas distributor channel (domain II) (see Fig.\ref{fig:CathodeConfigurations}). Therefore, a piece of the GDL is considered, which is partly covered by the land (shaded area in Fig.\ref{fig:GDL-PoreNetwork}), an area open for gas and water flow out of the GDL and a line of interface pores (red area in Fig.\ref{fig:GDL-PoreNetwork}), where the channel is opening. For simplicity reasons, the land surface is assumed to be hydrophobic but the surface in the channels is hydrophilic. Similar to previous numerical and experimental investigations in the literature \cite{bazylak2008dynamic,yu2018liquid,alink2013modeling}, the water passes through the GDL network (domain I) following preferential paths and leaves the GDL at a few breakthrough locations. As inlet boundary condition a mass flux corresponding to the water production at a current density of approx. $2A/cm^2$ is chosen. At the outlet a constant pressure with atmospheric conditions is set. Since the considered GDL sample is very small, only one breakthrough location is formed. The water saturation (non-wetting phase) in the network (domain I) before the breakthrough is shown in Fig.\ref{fig:GDL-breakthrough}.
\newline
The setup is designed such that this breakthrough location lays on the line of interacting pores (red line in Fig.\ref{fig:GDL-PoreNetwork}), where the hydrophobic network pores (domain I) are in contact with the hydrophilic gas distributor surface (domain II). The resulting saturation fluctuations in that pore are presented in Fig.\ref{fig:GDL-saturation}. 
\newline
The fluctuating behavior is similar to the one observed in the basic test scenarios. However, at the refilling of the pore, the water saturation curve shows a convex behavior which results from the highly heterogeneous network structure. While in the basic setup (Fig.\ref{fig:TopPores-saturations}), the interface pore is refilled by a few neighboring pores of the same size, the refilling neighboring pores in the GDL network vary in size. Also, as it can be seen in Fig.\ref{fig:GDL-breakthrough}, there are a lot more pores connected to each other (coordination number between 1 and 50 compared to coordination number between 2 and 4 in the basic test scenarios) than in the basic, structured network. The water pressure (non-wetting phase) difference to an empty interface pore is high, resulting in high fluxes through the connecting throats. Meanwhile, the smaller pores connected to the interface pore, which have a higher local capillary pressure, are emptied (or the water saturation is reduced significantly, such that the local capillary pressure drops) and the flux in the interface pore is reduced since the neighboring pores need to be refilled from the surrounding network (domain I) as well. Because the water can be distributed on the hydrophilic channel walls (domain II), the pressure does not build up during the emptying process. Therefore, the water saturation drops quickly after the threshold $s_{liq} = 0.52$ is reached. 
\newline
The high frequency of the oscillations results from the small pore sizes and high inlet flux of water for the small sample. \\
The considered GDL sample does not represent a representative volume of the GDL, it only shows the specific behavior at a certain breakthrough location. For representative studies, much larger GDL samples need to be considered. This is not the scope of this work, where we focus on the presentation of a concept to model the interaction of a hydrophobic porous medium (domain I) with a hydrophilic domain (domain II). However, the presented concept can be used for these kind of investigations and is the basis for further analysis of the hydrophobic-hydrophilic interactions between GDL and gas distributor structures.
\section{Final remarks}
\subsection{Discussion}
Especially for modeling the processes at the interface between GDL and gas distributor in a PEM fuel cell, we need to capture the effects of the different wetting behavior of the materials. The fluids (water and air) passing through the hydrophobic GDL (domain I) interact with the hydrophilic surface of the gas distributor channels (domain II) (see Fig.\ref{fig:CathodeConfigurations}). The presented concept considers the interaction at the contact line between GDL and channel wall and we focus on the water outtake from the hydrophobic GDL.\\
To develop a model that allows two phase flow in a hydrophobic porous medium (domain I) interacting with a hydrophilic structure (domain II), the following assumptions are used:
\begin{itemize}
	\item the GDL can be represented by a pore-network model,
	\item in the gas channel, the water is removed efficiently from the interface between the hydrophobic porous domain (GDL) (domain I) and the hydrophilic channel-land structure (domain II) such that the water saturation remains close to zero and the resulting local capillary pressure equals the non-wetting (gas) phase pressure at the interface,
	\item local saturation thresholds can be defined to identify the time frame of water interaction across the interface between the hydrophobic porous domain (GDL) (domain I) and the hydrophilic channel-land structure (domain II),
	\item these saturation thresholds are sufficiently independent of the connectivity of the boundary pore. Form the description of the pore and throat structure, analytic threshold formulations can be derived.
\end{itemize} 
With these assumptions a model is built that captures the main processes at the mixed-wet interface on the pore-scale. It allows to include the interface effects in a rather simple representation of the porous material (domain I) and helps to understand the ongoing processes and their interactive behavior. The developed model captures the effects of the mixed-wet interface not only on the bounding pores but also on the neighbors and the whole hydrophobic network flow (domain I). Therefore, the implementation is able to capture the dominating water flow behavior at the interface between a hydrophobic, porous structure (domain I) and a hydrophilic open flow (domain II) as it occurs at the interface between the cathode GDL and the gas distributor in a PEM fuel cell as shown in Fig.\ref{fig:CathodeConfigurations}.
\subsection{Outlook}
\begin{figure}[h!]
    \centering
	\includegraphics[width=0.8 \textwidth]{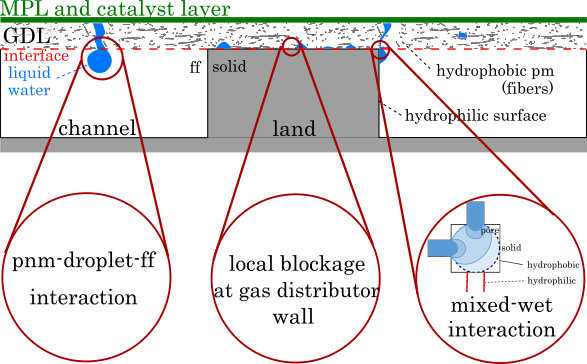}
	\caption{Different configurations occurring at the interface between GDL and gas distributor. In this work, an approach to capture the mixed-wet interactions is presented. This concept will be extended by an approach to capture the interaction between the GDL, modeled with a pore-network model, occurring droplets and the free flow (ff) in the gas channels, and by an approach to capture the interaction with the hydrophilic walls in regions where the water is trapped below the land.}
	\label{fig:InterfaceConfigurations}
\end{figure}

The processes are captured on the pore-scale using a pore-network model for the representation of the hydrophobic porous domain (domain I).\\
There are a few assumptions that had to be chosen such as the saturation thresholds defining the start and end of the fluid interaction. To overcome these assumptions further experimental and numerical investigations are necessary.\\
For a better understanding of the ongoing processes, experimental investigations using a microfluidic PDMS model with mixed-wet surface properties could be used.\\
The presented model is created to model the interaction between the GDL and the gas distributor in a PEM fuel cell. However, for further applications, the flexibility of the model allows extensions to different pore shapes and orientations, and the analysis of their influence. \\
The presented concept captures a part of the interface between GDL and gas distributor, shown by the left red circle in Fig.\ref{fig:CathodeConfigurations}. To investigate the performance of a polymer electrolyte membrane fuel cell (PEM FC) and to improve the water management in the cell, the presented mixed-wet interface concept shall be included in a model for GDL and gas distributor interaction representing the different occurring interface configurations as presented in Fig.\ref{fig:InterfaceConfigurations} (at the land parts, free flow interaction in the channel, mixed-wet interaction at the channel wall).
\section*{Acknowledgements}
We thank the Robert Bosch GmbH for the financial and professional support and the Deutsche Forschungsgemeinschaft (DFG, German Research Foundation) for supporting this work by funding SFB 1313, Project Number 327154368.

\bibliographystyle{ieeetr}
\bibliography{hydrophobic-hydrophilic-interaction} 

\end{document}